\documentclass[twocolumn]{article}
\usepackage{graphicx}
\usepackage{amscd}
\usepackage{amsfonts}
\usepackage{amssymb}
\usepackage{amsmath}
\begin{document}
\baselineskip=12pt
\title{Complete Gluonic Phase in Two-Flavor Color Superconductivity
\footnote{The project supported by the National Natural Science
Foundation of China under Grant No.10475113.}}
\author{Xiao-Ming Wang\footnote{E-mail:xiaomingw@mails.gucas.ac.cn} \ and Bang-Rong Zhou\footnote{E-mail:zhoubr@gucas.ac.cn}\\
{\small College of Physical Sciences, Graduate University of Chinese Academy of Sciences}\\
{\small Beijing 100049, China}}
\date{}
\maketitle
\begin{abstract}
We study a typical complete gluonic phase (LGP) in two-flavor color
superconductivity (2SC) by calculating the essential cubic and
quartic interfering term between the gluonic condensates $\langle
A^{(8)}_{z}\rangle$ and $\langle A^{(6)}_{z}\rangle$ with a gauged
NJL model. It is proven that the coefficient of the cubic
interfering term and the vacuum contributions of the cubic and
quartic interfering term are all equal to zeroes. The coefficients
of the quartic interfering term and the $\langle
A^{(6)}_{z}\rangle$'s quartic self-interaction term at stationary
points of LOFF phase are calculated. Comparisons among the effective
potentials of LGP, g2SC and LOFF phase indicate that LGP could be
the genuine ground state of 2SC for some reasonable parameters.
\end{abstract}
\maketitle \indent Since 1998, color superconductivity (CSC) in
dense quark matter has been a popular research topic. It has not
only important theoretical significance for understanding the phase
structure of Quantum Chromodynamics but also direct practical
applications to exploring the inner cores of compact stars. In
recent years, discovery of the chromomagnetic instability(CMI)
\cite{CMI} triggers another research tide, and many works have been
done to cure the CMI problem.\cite{GP3,GP31,GP32}.  In all the
present CMI-resolving schemes, the gluonic phase (GP) is naturally
of much interest \cite{GP,gsm}. However, the gluonic phase discussed
so far \cite{GP,GP3,GP31,GP32} can not be called complete gluonic
phase, because the gluonic condensate $\langle A^{(8)}_{z}\rangle$
was not included in the previous calculations. Since the gluon field
$A^{(8)}_{z}$ itself has also the problem of CMI, the genuine ground
state must contain the parameter $\langle A^{(8)}_{z}\rangle$. The
authors of the paper \cite{GP2} put forward some genuine complete
gluonic phase schemes including $\langle A^{(8)}_{z}\rangle$ (so
called phase A-phase F in Ref.\cite{GP2}). But up to now, no
detailed research has been done on their stability. \\
\indent It should be indicated that introduction of$ \langle
A^{(8)}_{z}\rangle $ in 2SC is equivalent to including the LOFF
phase extensively studied so far, with the correspondence that the
parameter of single plane wave LOFF state $ q \rightarrow
\frac{g}{2\sqrt{3}}\langle A^{(8)}_{z}\rangle$. Since the effective
potential of LOFF phase has been calculated precisely and some
simple analytic expressions for most of the other GP's parameters
have been obtained \cite{GP}, so  calculating the interfering terms
between $\langle A^{(8)}_{z}\rangle$ and the other parameters of the
GP becomes the key to derive the effective
potential of a complete gluonic phase.\\
\indent As stated above, introduction of the $\langle A_z^8\rangle$
is equivalent to inclusion of LOFF phase. However, in weak coupling
region, LOFF phase might well be the ground state of 2SC and if that
is true, then introduction of the other gluonic condensates will be
unnecessary, because in this case LOFF phase itself can cure the CMI
problem \cite{LOFF}. Otherwise, it will also be necessary to include
the gluonic condensates other than $\langle A_z^{(8)}\rangle$ . On
the other hand, in the intermediate and strong coupling region, it
has been proven that the ground state of 2SC can not be LOFF phase
\cite{LOFF2}, hence the left alternative could only be the complete
gluonic phase. For exploring such a phase, besides $\langle
A_z^8\rangle$, we must at least include $\langle A_z^{(6)}\rangle$
which is the remaining non-zero component of a complex
$SU_c(2)$-doublet composed of the gluonic condensates from
$\vec{A}^{(4)}-\vec{A}^{(7)}$ in the unitary gauge and after an
$O(3)$ rotation (similarly, $\langle A_z^8\rangle$ is the remainder
of $\langle \vec{A}^8\rangle$ after an $O(3)$ rotation), here the
$SU_c(2)$ is the remaining symmetry after the $SU_c(3)$ is broken by
the two-flavor diquark condensates. The other condensates from the
components $\langle \vec{A}^{(i)}\rangle(i=1,2,3)$, though possible
in principle,
will not be considered in present model. \\
\indent In this way, we will explore a typical edition of the
complete gluonic  phase (abbreviated as LGP, implying that the LOFF
phase with $\langle A_z^{(8)}\rangle$ and the GP
 with $\langle A_z^{(6)}\rangle$ coexist). Let $T_{Bq}$ and $\lambda_{Bq}$ denote respectively
 the coefficient of the cubic and quartic interfering term between $\langle A^{(8)}_{z}\rangle$
 and   $B=\langle A^{(6)}_{z}\rangle$,
 and $\lambda_{B}$ denote the coefficient of the quartic self-interaction term of $B$. We will
 only calculate
 the values of these coefficients and the LGP effective potential at the LOFF phase's ground state
 so as to examine the essential character of LGP. This is because the
 interfering terms
 between $\langle A^{(8)}_{z}\rangle$ and $B=\langle A^{(6)}_{z}\rangle$
 haven't simple analytic expressions, so it is difficult to find out the solutions of the
     neutral condition and the gap equations and to fix the ground state's parameters of LGP.
     Hence, we have to adopt a
     conventional practice instead, i.e. takeing the ground state's parameters of other
     approximate phase
     as the ones of present model. It is indicated that the ground state's parameters of LOFF phase
     are close to the ones of  LGP.  In fact, the main part in the effective potential of LGP
     comes from 2SC+LOFF phase, thus the ground state's parameters of LGP will be mainly determined by the 2SC+LOFF
     sector in  LGP. This point has been verified indirectly by the results in Refs. \cite{GP3,GP31}. For example,
     the parameters of the gluonic cylindrical phase II's ground state shown in \textbf{Fig.1} and \textbf{Fig.2 }
     of Ref. \cite{GP31} is surprisingly similar to those of the LOFF phase'ground state obtained by us and the
     small difference between the both may be attributed to different approximations to the effective potentials.
     Thus it is
     certainly rational to calculate the effective potential of LGP at the
     stationary points of LOFF phase. In addition,
     we can see from Refs.\cite{GP3,GP31} that the gluonic phase only gives  no more than ten percent contribution
     to the total effective potential of 2SC +GP, hence in the Ginzburg-Landzu approach, the effective potential
     of LGP will be expanded in the gluonic condensates only up to the fourth order. \\
     \indent We will use a gauged NJL model describing two-flavor color superconductivity at zero
     temperature. When there exist diquark condensates, we
     introduce the Nambu-Gor'kov bispinor field
     \begin{equation}
     \Psi= \frac{1}{\sqrt{2}}\left(%
     \begin{array}{c}
     q \\
     q^{C} \\
     \end{array}%
     \right)
     \end{equation}
     with the quark field
     $q\equiv q_{i\alpha}$ which is a four-component Dirac spinor
     with the
     flavor($i=u,d$) and the color ($\alpha=r,g,b$)
     indices and the charge conjugated field defined
     by
     $q^{C}=C\overline{q}^{T}$.
     Thus the Lagrangian density can be written as
     \begin{equation}\label{ld}
     \mathcal{L}=\overline{\Psi}\mathcal{S}^{-1}\Psi,
     \end{equation}
     where the inverse propagator of the field $\Psi$
     \begin{equation}\label{a}
     \mathcal{S}^{-1}=\left(%
     \begin{array}{cc}
     [G^{+} _{0}]^{-1}& \Delta^{-1} \\
     \Delta^{+} & [G^{-}_{0}]^{-1} \\
     \end{array}%
     \right)
     \end{equation}
     with
     \begin{eqnarray}\label{b}
     [G^{+}_{0}]^{-1}(K)&=&(k_{0}+\overline{\mu}-\delta\mu\tau_{3}-\mu_{8}\mathbf{1}_{b})\gamma^{0}
     \nonumber\\&&-\vec{\gamma}\cdot\vec{k},\\
     \label{c}[G^{-}_{0}]^{-1}(K)&=&(k_{0}-\overline{\mu}+\delta\mu\tau_{3}+\mu_{8}\mathbf{1}_{b})\gamma^{0}\nonumber\\
     &&-\vec{\gamma}\cdot\vec{k},\\
     \label{d}\Delta^{-}&=&-\imath\varepsilon\varepsilon^{b}\gamma_{5}\Delta,\nonumber\\
     \Delta^{+}&=&\gamma^{0}(\Delta^{-})^{\dagger}\gamma^{0}=-\imath\varepsilon\varepsilon^{b}\gamma_{5}\Delta
     \end{eqnarray}
     The denotations $\varepsilon$, $\varepsilon^{b}$,
     $\overline{\mu}$, $\delta\mu$, $\mu_{8}$, etc. in formulas (\ref{b})-(\ref{d}) are the same
     as the ones used in Ref. \cite{GP2}.\\
     \indent The propagator of $\Psi$ is given by
     \begin{equation}
     \left(%
     \begin{array}{cc}
     G^{+} & \Xi^{-}\\
     \Xi^{+}& G^{-} \\
     \end{array}%
     \right)
     \end{equation}
     with
     $G^{\pm}=([G^{\pm}_{0}]^{-1}-\Delta^{\mp}G^{\mp}_{0}\Delta^{\pm})^{-1}$,
     $\Xi^{\pm}=-G^{\mp}_{0}\Delta^{\pm}G^{\pm}$.\\
     \indent In order to cure CMI problem, we must extend
     $S^{-1}$ to including the VEVs of relevant gluon
      fields. The effective potential of LGP in the Ginzburg-Landau approach can be generally expressed by
      \cite{GP2}
      \begin{eqnarray}\label{GP}
     V_{LGP}&=&\frac{\Delta^{2}}{4G_{D}}-\frac{1}{2}\int\frac{d^4k}{i(2\pi)^{4}}Tr\mathcal{S}^{-1}_{g}\nonumber\\
     &&+\frac{1}{2}\int\frac{d^4k}{i(2\pi)^{4}}Tr\mathcal{S}^{-1}_{g}|_{\mu=\mu_{e}=\Delta=0}\nonumber\\
     &=&V_{\Delta}+\Sigma^{\infty}_{n=1}\frac{(-1)^{n}}{2n}\int\frac{d^4k}{i(2\pi)^{4}}Tr{(\mathcal{SM}_{g})}^{n}\nonumber\\
     &&+\frac{1}{2}\int\frac{d^4k}{i(2\pi)^{4}}Tr\mathcal{S}^{-1}_{g}|_{\mu=\mu_{e}=\Delta=0},
     \end{eqnarray}
     where $V_{\Delta}$ represents the effective potential of 2SC and
     $\mathcal{S}^{-1}_{g}=S^{-1}+\mathcal{M}_{g}$
     with $\mathcal{M}_{g}$ denoting the in- verse propagator of relevant gluon fields.
     In present model, as has been stated above, only the two most
     important VEVs of gluon fields
     $\langle A^{(8)}_{z}\rangle$ and $B=\langle A^{(6)}_{z}\rangle$ will
     be included in $\mathcal{M}_{g}$. We will also use a
     simplifying treatment, i.e. setting $\mu_8=0$, as was done in
     other papers. On these grounds, through a lengthy calculation of relevant one-loop diagrams from the fermion
     determinant, we have got the expressions of $T_{Bq}$ and $\lambda_{Bq}$ in the form
     of products of fermion propagators. The the last term in
     (\ref{GP}) contains the vacuum contributions of the cubic and
     quartic interfering term between $\langle A^{(8)}_{z}\rangle$ and
     $B=\langle A^{(6)}_{z}\rangle$ which, however, are eventually
     proven to be equal to zeroes after a hard and careful calculation.
     As a result, the derived effective potential becomes
     \begin{eqnarray} \label{LGP}
     V_{LGP}&=&V_{\Delta}+V_{LOFF}+M^{2}_{B}B^{2}/2+\lambda_{B}B^{4}/4\nonumber\\
     &&+T_{Bq}B^{2}q+\lambda_{Bq}B^{2}q^{2}/2,
     \end{eqnarray}
     where $V_{LOFF}$ is the effective potential of LOFF phase,
     $M_B^2$ is the squared Meissner mass of the gluonic field
     $A_z^{(6)}$.
     Furthermore, it has been proven that the coefficient of the cubic interfering term is
     equal to zero, i.e. $T_{Bq}=0$. Hence the extreme value condition about parameter
     $B$ derived from (\ref{LGP}) becomes
     \begin{equation}\label{bev}
     \partial V_{LGP}/\partial B=M^{2}_{B}B+\lambda_{B}B^{3}+\lambda_{Bq}Bq^{2}=0.
     \end{equation}
     The solution of (\ref{bev}) is  $B = 0$  and  \  \
     $ B = B_1=\sqrt{(-M^{2}_{B}- \lambda_{Bq} q^{2})/ \lambda_{B}}$.
     It is seen from (9) that $B=0$ just corresponds to the 2SC+ LOFF
     phase. By analyzing the second-order derivative of $V_{LGP}$ over $B$, we find that
     if $\lambda_{Bq}<-M^2_B/q^2$, then $B=B_1$ and
     $B=0$ will respectively be a minimal and a maximal point of
     $V_{LGP}$ about $B$  and this means that LGP could have lower energy than  2SC + LOFF
     phase in this case.
     Now considering only the situation of $B\neq0$ and substituting $B_1$
     into the effective potential (\ref{LGP}), we may change $V_{LGP}$ into $\bar{V}_{LGP}$,
     where
     \begin{eqnarray} \label{LGP2}
     \bar{V}_{LGP}
     &=&V_{\Delta}+V_{LOFF}-\lambda_{B}B^{4}/4\nonumber \\
     &=&V_{\Delta}+V_{LOFF}\nonumber\\&&-(-M^{2}_{B}-\lambda_{Bq}q^{2})^{2}/4\lambda_{B}.
     \end{eqnarray}
     $\bar{V}_{LGP}$ is the effective potential of LGP at its
     non-zero stationary point about the parameter $B$ and will become our
     starting point to make a comparison among the effective potentials of LGP, g2SC and LOFF
     phase. In fact, from (\ref{LGP2}), we can see that if $\lambda_{B}>0$, then$\bar{V}_{LGP}$ could be lower than
     $V_{\Delta}+V_{LOFF}$.\\
     \indent The above qualitative analysis indicates that for making a comparison among the effective potentials, the key point is to
     calculate $\lambda_{Bq}$ and $\lambda_{B}$. We find that in doing so, the most difficult task is to
     remove some superficial singularities. This is because after the energy summing,
     in the denominators of the integrands there will appear the factors similar to
     $a=p^0-\sqrt{(|\vec{k}|+\bar{\mu})^2+\Delta^2}+$$\sqrt{(|\vec{k}|+|\vec{p}|+\bar{\mu})^2
     +\Delta^2}$ which could become zero when the gluons' four-momentum $(p^0,\vec{p})\rightarrow 0$
     and this will lead to singularity in the integral over the quark momentum $|\vec{k}|$.
     To remove such singularities, we invent a symbol calculating method.
     First denote these pole factors respectively by the
     symbols $a$,$b$,$c$,$d$,$e$,$f$ and express an
     integrand generally in the form $f(a,b,c,d,e,f)/abcdef$,
     then by means of some found interrelations between
     these symbols, through lengthy calculations (since the quark propagators have
     large number -four- and very complicated forms), we have
     ultimately proven that all the pole factors in the
     denominators can be canceled out one by one and the superficial singularities are
     removed.

     In our work, another key calculation is to fix the parameters of the
     ground state of LOFF phase.  Although this calculation has been worked out \cite{LOFF2}, the detail of calculation
     wasn't given.  In fact we have conducted this calculation independently and
     obtained the exactly same results as those shown in Ref. \cite{LOFF2}. This lays a solid
     foundation for our following move. The last difficult task is to do the momentum integrations involving
     tedious numerical computations. The final results are shown
     in Fig.1, Fig.2 and Fig.3\\
     \begin{figure}
     {\centering
     \includegraphics[width=0.5\textwidth]{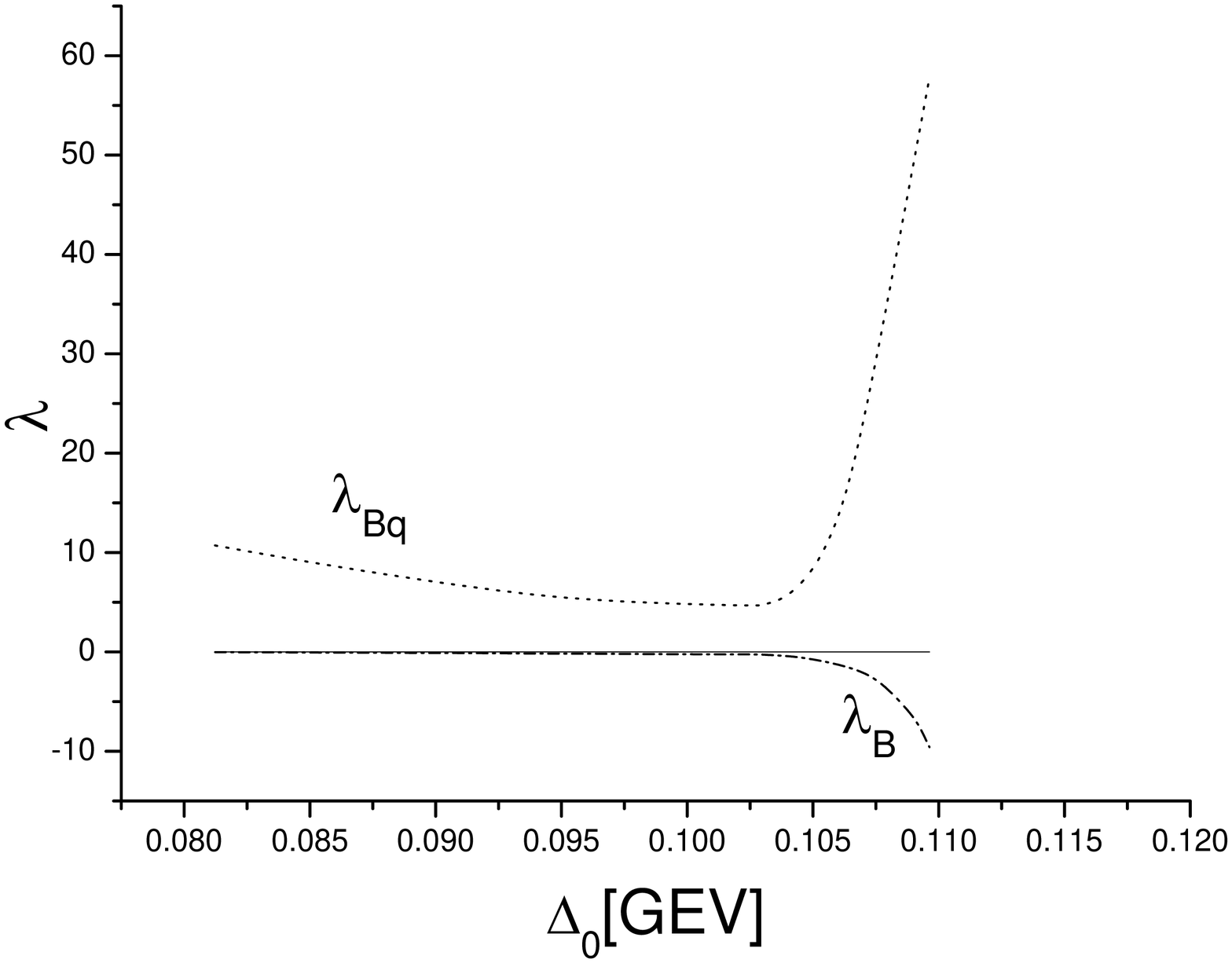}}
     {\small Fig.1: Varying of $\lambda_{Bq}$ and $\lambda_{B}$ as LOFF phase's stationary
     point when $\delta\mu>\Delta$ ($\mu=0.4$ GeV, $\Lambda=0.6533$ GeV).}\label{figure1}
     \end{figure}
     \begin{figure}
     {\centering
     \includegraphics[width=0.5\textwidth]{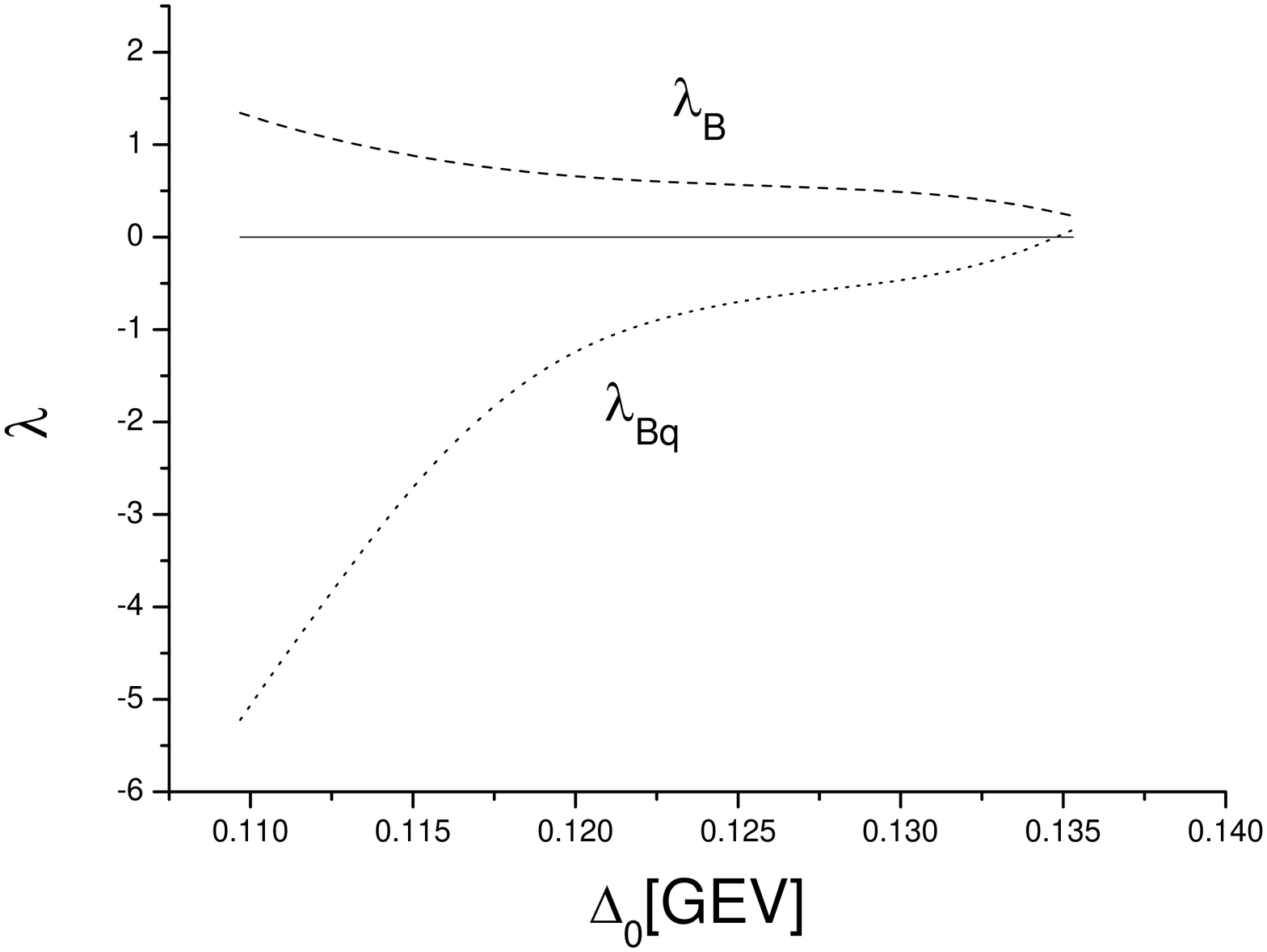}}
     {\small Fig.2: Varying of $\lambda_{Bq}$ and $\lambda_{B}$ as LOFF phase's stationary point when
     $\delta\mu<\Delta$ ($\mu=0.4$ GeV, $\Lambda=0.6533$ GeV).}\label{figure2}
     \end{figure}
     \indent All the x-axes of the three figures are $\Delta_{0}$ which is the gap of the 2SC
     with $\delta\mu=0$ and one-one corresponding to the coupling strength $G_D$ of
     diquark condensates.
     For a fixed $\Delta_0 \,(G_D)$ we may find out a stationary
     point of LOFF phase (when $\mu$ is given), and in the
     meantime, a respective definite value of $\lambda_{Bq}$,
     $\lambda_{B}$ and the effective potentials, while
     different $\Delta_0$ may correspond to different
     parameters $\delta\mu$ and $\Delta$. Based on these data, we will be able to compare
     the least energy states of LGP, g2SC and LOFF phase.\\
     \indent From Fig.1, it may be seen that at the stationary point of LOFF phase
     when $\delta\mu>\Delta$, we have $\lambda_{Bq}>0$ (dot line)
     and $\lambda_{B}<0$ (dot-dash line). The value of $\lambda_{B}$ is very small in the
     region $\Delta_{0}<0.10269$ GeV ( e.g. $\Delta_{0}=0.081225$ GeV, $\lambda_{B}\approx0.018$
     and $\Delta_{0}=0.10269$ GeV, $\lambda_{B}\approx0.24$). Near the critical
     point $\delta\mu=\Delta$ ($\Delta_{0} =$ 0.109683 GeV), the absolute values
     of $\lambda_{Bq}$ and $\lambda_{B}$ become very large, this attributes to some
     terms containing the $\delta$ function  $\delta \left(\delta\mu-\sqrt{(|\vec{k}|-\bar{\mu})^{2}+\Delta^2}\right)$  \\
     in $\lambda_{Bq}$
     and $\lambda_{B}$ which become very large near the critical point after the momentum integral.
     In view of (\ref{LGP2}), the above result shows that when $\delta\mu>\Delta$, the effective potential of LGP is higher than that of LOFF phase.\\
     \indent From Fig.2, we can see that at the stationary point of LOFF phase
     when $\delta\mu \leq \Delta$ ($0.109683$ GeV $\leq \Delta_{0} \lesssim0.137$ GeV), the
     result is that $\lambda_{Bq}<0 \;(0.109683$ GeV$<\Delta_{0}\lesssim0.134$ GeV) (dot line)
     and $\lambda_{Bq}>0(0.134$ GeV $\lesssim\Delta_{0}\lesssim0.137$ GeV) and
     $\lambda_{B}>0$ (dash line).\\
     \begin{figure}
     {\centering
     \includegraphics[width=0.5\textwidth]{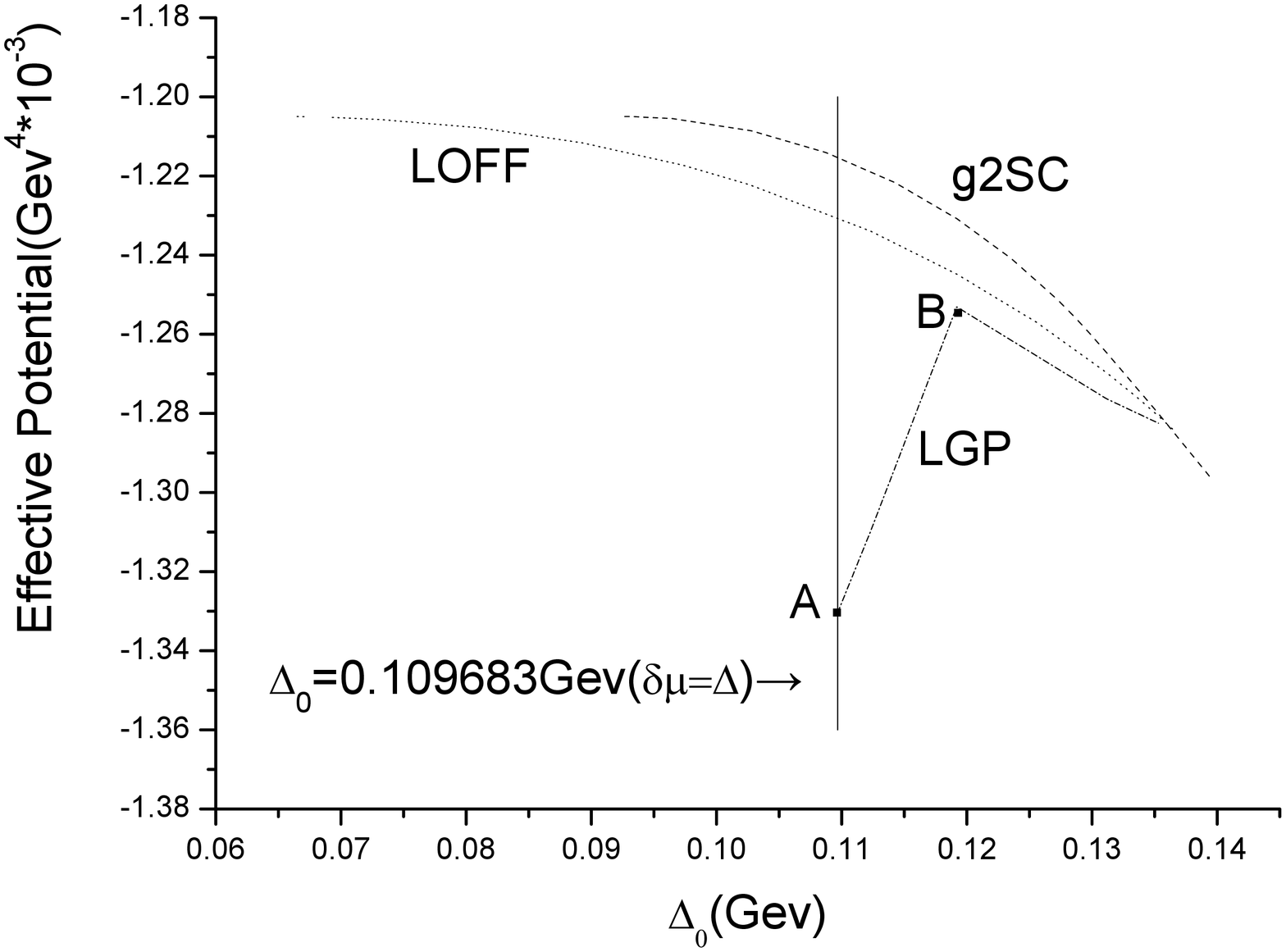}}
     {\small Fig.3: Effective potential of LGP, g2SC and LOFF phase at LOFF's stationary points ($\mu=0.4$ Gev, $\Lambda=0.6533$ Gev).}\label{figure3}
     \end{figure}
     \indent In Fig.3, we have drawn the effective potentials of LGP, g2SC and  LOFF phase.
     Here we only give the effective potential of LGP when $\delta\mu<\Delta$ and
     omit the one when $\delta\mu>\Delta$, since in the latter case LGP has higher
     effective potential than LOFF phase at the stationary point of LOFF phase.
     From Fig.3, we can see that the effective potential of LOFF phase (dot line) is lower than that of the g2SC phase(dash line) in
     the whole LOFF phase window $0.065$ GeV $<\Delta_{0}<0.137$ GeV. However, LGP (dot-dash line) has lower
     effective potential than LOFF phase at stationary point of LOFF phase when $\delta\mu<\Delta$ ($0.10968$ GeV $<\Delta_{0}<0.137$ GeV).
     Hence, we may conclude that in this region,
     LGP could be the ground state of the system. \\
     \indent In Fig.3, the abnormal part in the effective potential of LGP, namely, from point A to point B,
     is caused by the sudden vanishing of the terms containing $\delta$
     function  $\delta\left(\delta\mu-\sqrt{(|\vec{k}|-\overline{\mu})^{2}+\Delta^2}\right)$.
     With regard to the situation at the LOFF phase's stationary point when
     $\delta\mu>\Delta$ i.e.
     in weak coupling region, the genuine ground state could be
     either LOFF state, or another phase when other parameters are
     considered. \\
     \indent The above study of LGP shows that the complete gluonic phase indeed possesses the
      character of the genuine ground state for some reasonable parameters. Although present
      work is preliminary and remains to be improved further, it is believed that the derived
      qualitative conclusion will not be changed essentially.
      
        \end{document}